\documentclass[12pt]{article}
\usepackage{a4}
\usepackage{latexsym} 
\usepackage{amssymb}  
\usepackage{graphicx}
\usepackage{epsfig}

\newcounter{zyxabstract}     
\newcounter{zyxrefers}        

\newcommand{\newabstract}
{\newpage\stepcounter{zyxabstract}\setcounter{equation}{0}
\setcounter{footnote}{0}}

\newcommand{\rlabel}[1]{\label{zyx\arabic{zyxabstract}#1}}
\newcommand{\rref}[1]{\ref{zyx\arabic{zyxabstract}#1}}

\renewenvironment{thebibliography}[1] 
{\section*{References}\setcounter{zyxrefers}{0}
\begin{list}
{[\arabic{zyxrefers}]}
{\usecounter{zyxrefers}\setlength{\parindent}{0cm}\setlength{\itemsep}{0cm}}

}
{\end{list}}
{\section*{References}\setcounter{zyxrefers}{0}
\begin{list}{[\arabic{zyxrefers}]}
{\usecounter{zyxrefers}\setlength{\parindent}{0cm}\setlength{\itemsep}{-1.5mm}}}
{\end{list}}

\renewcommand{\bibitem}[1]{\item\rlabel{y#1}}
\renewcommand{\cite}[1]{[\rref{y#1}]}      

\newcommand{\Opc}{{\mathcal{O}}(p^4)}

\begin{document}
\begin{titlepage}

\begin{flushleft} 
\includegraphics*[height=2.2cm]{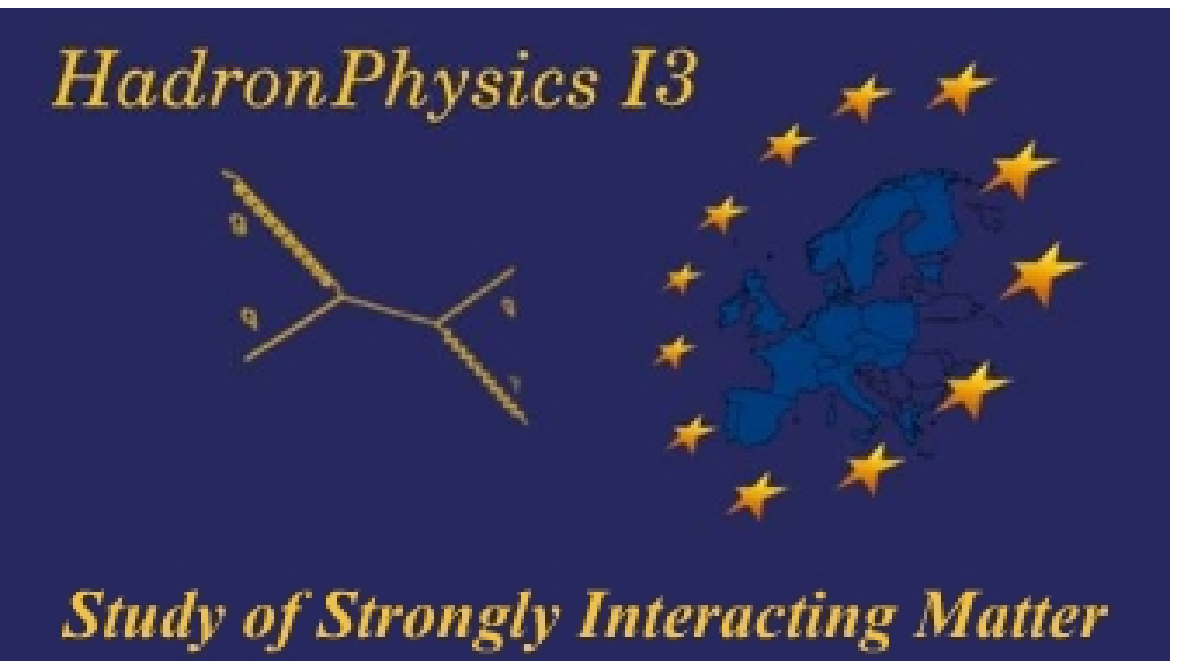}
\end{flushleft}

\vspace{-2.5cm}

\begin{flushright} 
{\small
HISKP-TH-06/32\\
FZJ-IKP-TH-2006-27\\
Internal Report, DESY Zeuthen 06-01\\
}
\end{flushright}

\vspace{0.6cm}

\begin{center}
{\huge\bf Lattice QCD,\\[1.5mm] Chiral Perturbation Theory \\[2.1mm]
and Hadron Phenomenology$^{*}$}
\\[1cm]
ECT* -- I3HP Workshop,
ECT*, Trento, Italy,
October 2 - 6, 2006\\[1cm]
{\bf Ulf-G. Mei\ss ner}$^{1,2}$ 
and {\bf Gerrit Schierholz}$^{3,4}$\\[0.3cm]
$^1${Universit\"at Bonn, Helmholtz-Institut f\"ur Strahlen- und Kernphysik (Theorie)\\ 
D-53115 Bonn, Germany}\\[0.3cm]
$^2${Forschungszentrum J\"ulich, Institut f\"ur Kernphysik (Theorie)\\ 
D-52425 J\"ulich, Germany}\\[0.3cm]
$^3${Deutsches Elektronensynchrotron DESY, D-22603 Hamburg, Germany}\\[0.3cm]
$^4${John von Neumann-Institut f\"ur Computing NIC\\
Deutsches Elektronen-Synchrotron DESY, D-15738 Zeuthen, Germany}\\[1cm]
{\large ABSTRACT}
\end{center}
These are the proceedings of the workshop on ``Lattice QCD, 
Chiral Perturbation Theory and Hadron Phenomenology''
held at the European Centre for Theoretical Studies in
Nuclear Physics and Related Areas from October 2 to 6, 2006.
The workshop concentrated on bringing together researchers
working in lattice QCD and chiral perturbation theory
with the aim of improving our understanding how hadron
properties can be calculated and analyzed from first principles.
Included are a short contribution per talk.

\vspace{0.3cm}

\noindent\rule{6cm}{0.3pt}\\
\footnotesize{$^*$ This workshop was funded by the ECT*
and by the EU Integrated Infrastructure Initiative Hadron
Physics Project under contract number RII3-CT-2004-506078.
Work supported in part by DFG (SFB/TR 16 ``Subnuclear Structure of
Matter'' and FOR 465 ``Forschergruppe Gitter-Hadronen-Ph\"anomenologie'')
and by BMBF (research grant 06BN411).}

\end{titlepage}

\section{Introduction}
As a result of developments in lattice field theory and computer
technology, the first full QCD calculations in the chiral regime are
becoming available now. An essential tool for extracting hadronic
quantities from lattice QCD is chiral perturbation theory.
Chiral perturbation theory is the effective field theory of QCD
and allows to analyze the quark mass dependence of observables
in a model-independent manner. Since the
technologies for large scale lattice calculations and for detailed chiral
perturbation calculations are each so different and so demanding, the same
theorists generally do not pursue both aspects. Hence, we considered
it to  be valuable to bring together practioners of both aspects of 
lattice calculations, and
deepen each others' appreciation of the issues involved, thus opening a
new era for confronting experiment with solutions of QCD from first
principles.

Thus, such a meeting was organized at the ECT* (Trento) from October
2-6, 2006, with financial support from the ECT* as well as the I3HP
networks N2 and N5 and by EURONS. The meeting had 36 participants whose names, 
institutes and email addresses
are listed below. 35 of them presented results in presentations of various
lengths. Ample time was left for discussions on the various talks each day.
In addition, there was also a special discussion session 
``Lattice meets chiral perturbation theory'' convened by Hartmut Wittig, 
in which the interplay between the various issues as seen from the two 
communities were addressed.
A short description of the contents of each talk and a list of the most relevant
references can be found below. We felt that this was more appropriate a 
framework than full-fledged proceedings.
Most results are or will soon be published and available on the archives,
so this way we can achieve speedy publication and avoid duplication of
results in the archives.

Below follows first the program, then the list of participants 
followed by the abstracts of the talks. The summary of the aforementioned
discussion session is given at the end of these proceedings. Most talks 
can also be obtained from the workshop website 

\centerline{{\tt http://www.ect.it/}}

\bigskip

We would like to thank the ECT*, in particular Serena Degli Avancini, 
for the excellent organization of the workshop and all participants 
for their valuable contributions. We believe that this was only the
first workshop of this kind and look forward to similar meetings in the
future.

\vspace{1.2cm}

\hfill Ulf-G. Mei\ss ner and Gerrit Schierholz

\vfill \eject

\section{Program}

\vspace{0.cm}

\begin{tabbing}
xx:xx \= A very very very long name \= \kill
{\bf Monday, October 2nd 2006}\\[3mm]
{\em Morning session,  chair: Evgeny Epelbaum}  \\
  9:45 \>    Ulf-G. Mei{\ss}ner /    \>  Opening Remarks \\
        \>    Gerrit Schierholz  \> \\
9:50\>        Luigi Scorzato (Trento)  \> 
   Dynamical simulations with twisted mass   \\
 10:25 \>
        Gerrit Schierholz \>
Lattice QCD simulations at small quark masses\\
\> (Hamburg) \\
11:00\>\> {\em Coffee}\\
11:30 \>
        Christian Lang (Graz)\>         
Hadrons for chirally improved fermions\\
12:15 \>
        Meifeng Lin (New York)  \>
Dynamical domain wall fermion simulations \\
\> \> and chiral perturbation theory\\
13:00\>\> {\em Lunch}\\
{\em Afternoon session, chair: Bugra Borasoy} \\
14:30 \> 
        Christopher Aubin  \>      
Applications of staggered ChPT\\
\> (New York)\\
15:15 \>
         Stephan D\"urr (Bern) $^\dagger$ \>      
 Are there alternatives to rooted staggered fermions? \\
15:50\>\> {\em Coffee}\\
15:15 \>
         Sinya Aoki (Tsukuba) \>
 ChPT and lattice QCD with Wilson-type quarks\\
 17:05\>\> {\em Discussions}\\  
18:00\>\>{\em End of Session}\\[5mm]
{\bf  Tuesday, October 3rd, 2006}\\[3mm]
{\it  Morning Session, chair: Meinulf G\"ockeler}\\
09:00 \>
        Oliver B\"ar (Berlin)\>   Lattice QCD with mixed actions: \\
\>\> Overlap fermions on a twisted mass sea\\
09:35\>
        Ulf-G. Mei{\ss}ner (Bonn)\>
Thoughts on chiral extrapolations for excited states \\
10:10 \>
        V\a'eronique Bernard \>    
 Chiral extrapolations of baryon properties\\ \> (Strasbourg)\\ 
10:45 \>\> {\em Coffee}\\
11:15\>
        Matthias Schindler (Mainz)  \> 
       Some topics in manifestly Lorentz-invariant CHPT\\
11:50\>
        Evgeny Epelbaum (J\"ulich)  \>  
        Chiral extrapolations in few-nucleon systems\\
12:25\>\> {\em Lunch}\\
{\em Afternoon session, chair: Christopher Aubin} \\
14:30\>
        Bugra Borasoy (Bonn) \> Nuclear lattice simulations\\
15:15\>
        Jos\a'e A. Oller (Murcia)\> 
 Chiral non-perturbative study of pseudoscalar self-energies \\ 
15:50 \>\> {\em Coffee}\\
16:20\>
       Hermann Krebs (J\"ulich)    \>   
       Scalar two-loop diagrams on the lattice \\
16:55\> Dina Alexandrou (Nicosia) \> $N$ and $N$ to $\Delta$  form
factors in lattice QCD\\
 17:40\>\> {\em Discussions}\\  
18:00\>\>{\em End of Session}\\[6mm]
{\bf Wednesday, October 4th, 2006}\\[3mm]
{\em Morning Session, chair: Akaki Rusetsky}\\
09:00\>
        David Richards  \>  Hadron
structure using DWF quarks on an asqtad sea \\  
\> (Newport News) \\
09:45\>
        Wolfram Schroers (Zeuthen) \> Distribution amplitudes from the lattice\\
10:20\>
        Dirk Br\"ommel (Hamburg)  \> Pion structure from
        lattice QCD \\
10:55\>\> {\em Coffee}\\
11:55\>
       Dirk Pleiter (Zeuthen) \>
       Nucleon structure functions and form factors\\
12:10\>
        Enno Scholz (Upton) \> 
       $B_K$ from dynamical domain wall fermion simulations\\
12:45\>\>{\em Lunch}\\
{\em Afternoon Session, chair: Meifeng Lin}\\
14:30\>
        Hartmut Wittig (Mainz)  \>  
        Low-energy constants for $\Delta S=1$ transitions \\
15:05\>
         Akaki Rusetsky (Bonn)  \> 
         The $\Delta$--resonance in a finite volume \\ 
15:40\>\> {\em Coffee}\\
16:10\>
        Meinulf G\"ockeler  \> 
        Lattice QCD on smaller and larger volumes \\
\> (Regensburg) \\
16:55\>
        Tobias Gail (Garching)  \>  
Utilizing covariant BChPT for chiral extrapolations \\
17:30\> \>{\em Discussions}\\
18:00\> \>{\em End of Session}\\[15mm]
{\bf Thursday, October 5th, 2006}\\[3mm]
{\em Morning Session, chair: Oliver B\"ar}\\
9:00\> Christoph Haefeli (Bern) \> Aspects of CHPT at large $m_s$\\
9:45\> Poul Damgaard  \> A chiral two-matrix theory
and the chiral Lagrangian \\
\> (Copenhagen) \\
10:30\>\> {\em Coffee}\\
11:00\>
        Jack Laiho (Batavia) \>  The $B \to D^* \ell \nu$  form
factor from  lattice QCD\\
11:45\>
        Michele Della Morte \> 
         Heavy quark effective theory on the lattice:\\
\> (Geneva) \> The b-quark mass including ${\cal O}(1/m_b)$\\
12:25\>\>{\em Lunch}\\
{\em Afternoon Session, chair: Keh-Fei Liu}\\
14:30\>
        Wolfgang Bietenholz    \>   
Overlap hypercube quarks in the $p$- and the $\epsilon$-regime\\
\> (Belin) \\
15:05\>
        \>{\em Discussion: Lattice meets chiral perturbation theory}~$^\ddagger$\\
       \> \>  Hartmut Wittig (convenor)\\
16:35\> \>{\em End of Session}\\[6mm]
\end{tabbing}

\vfill\eject

\begin{tabbing}
xx:xx \= A very very very long name \= \kill
{\bf Friday, October 6th, 2006}\\[3mm]
{\em Morning Session, chair: Christian Lang}\\
09:00\>
        Keh-Fei Liu (Kentucky)   \>       
$\sigma(600)$ as a tetraquark mesonium and\\ 
\>\>the pattern of scalar mesons\\
09:35\>
        Marina Dorati (Pavia)    \>      
Moments of generalized parton distribution functions\\
10:10\>
        Andreas Sch\"afer  \>  
 ChPT for generalized parton distributions (GPDs)\\
\> (Regensburg)\\
10:20\> \>{\em Coffee}\\
11:15\>
        Volker Weinberg (Zeuthen)     \>    
The QCD vacuum as seen by overlap fermions \\ 
11:50\>
        Pietro Faccioli (Trento)  \>  
    The chiral regime of QCD in the instanton liquid model \\
12:25\>
        Christian Lang (Graz)  \> Farewell\\
12:30\> \>{\em Lunch and End of Workshop}\\[15mm]
\end{tabbing}

\vfill

\rule{9.7cm}{0.5pt}

$^\dagger$ no abstract was provided

$^\ddagger$ summary given at the end of these mini-proceedings
\eject

\section{Participants and their email}


\begin{tabbing}
A very long namexxxxx\=a very long institutexxxxxx\=email\kill
C. Alexandrou\> Univ. of Cyprus\> alexand@ucy.ac.cy\\
S. Aoki\> Univ. Tsukuba\> saoki@het.ph.tsukuba.ac.jp\\
C. Aubin\>Columbia Univ.\> caubin@phys.columbia.edu\\
O. B\"ar\> Humboldt Univ. Berlin\> obaer@physik.hu-berlin.de\\
V. Bernard \>ULP Strasbourg\> bernard@lpt6.u-strasbg.fr\\
W. Bietenholz\>Humboldt Univ.\>bietenho@physik.hu-berlin.de\\
B. Borasoy\> Univ. Bonn\>borasoy@itkp.uni-bonn.de\\
D. Br\"ommel\> DESY \> dirk.broemmel@desy.de\\
P. H. Damgaard\> Niels Bohr Institute\> phdamg@nbi.dk\\
M. Della Morte\> CERN \>  dellamor@mail.cern.ch \\
M. Dorati\> Univ. Pavia\>marina.dorati@pv.infn.it\\
S. D\"urr\> Bern Univ.\>durr@itp.unibe.ch\\
E. Epelbaum\>FZ J\"ulich \& Univ. Bonn\>epelbaum@fz-juelich.de\\
P. Faccioli\> Univ. Trento\> faccioli@science.unitn.it\\
T. Gail\> TU M\"unchen\>tgail@ph.tum.de\\
M. G\"ockeler\> Univ. Regensburg\>meinulf.goeckeler@physik.uni-regensburg.de\\
C. Haefeli\> Bern Univ.\> haefeli@itp.unibe.ch\\
K. Koller\> LMU M\"unchen \> karl.koller@lrz.uni-muenchen.de\\
H. Krebs\> FZ J\"ulich \& Univ. Bonn\>hkrebs@itkp.uni-bonn.de\\
J. Laiho\> Fermilab\>jlaiho@fnal.gov\\
C. B. Lang\> Univ. Graz\> christian.lang@uni-graz.at\\
M. Lin\>Columbia Univ.\> mflin@physics.columbia.edu\\
K.-F. Liu\>Kentucky Univ.\>liu@pa.uky.edu\\
U.-G. Mei{\ss}ner\> Univ. Bonn \& FZ J\"ulich\> meissner@itkp.uni-bonn.de\\
J. Oller\> Univ. Murcia\> oller@um.es\\
D. Pleiter\> DESY \> Dirk.Pleiter@desy.de\\
D. Richards\> Jefferson Lab\> dgr@jlab.org\\
A. Rusetsky\> Univ. Bonn\>rusetsky@itkp.uni-bonn.de\\
A. Sch\"afer\> Univ. Regensburg\> andreas.schaefer@physik.uni-regensburg.de\\
G. Schierholz \> DESY\>  Gerrit.Schierholz@desy.de\\
M. Schindler\> Univ. Mainz\> schindle@kph.uni-mainz.de\\
E. E. Scholz\> Brookhaven National Lab\> scholzee@quark.phy.bnl.gov\\
W. Schroers\> DESY \> Wolfram.Schroers@desy.de\\
L. Scorzato\>ECT Trento\> scorzato@ect.it\\
V. Weinberg\> DESY \> Volker.Weinberg@desy.de\\
H. Wittig\> Univ. Mainz\> wittig@kph.uni-mainz.de\\
\end{tabbing}

\newabstract 

\begin{center}
{\large\bf Results from Lattice QCD simulations with a twisted mass}\\[0.5cm]
{\bf Luigi Scorzato}$^1$, on behalf of the ETM Collaboration\\[0.3cm]
$^1$ECT*, strada delle tabarelle 286. 38050 - Villazzano (TN). Italy\\
\end{center}

We report on first results of an ongoing effort to simulate lattice QCD with 
two degenerate flavours of quarks by means of the twisted mass formulation 
tuned to maximal twist. We obtain pseudo-scalar masses below 300 MeV on 
volumes with spatial size larger than 2 fm at values of the lattice spacing 
similar or smaller than 0.1 fm. \cite{Jansen:2006rf}.
We present first comparison with ChPT including Finite Size Effects.
With the help of Wilson-ChPT we give evidence that $O(a^2)$ 
lattice artifacts are under control.
Additionally, exploratory results for the case of $N_f=2+1+1$ flavours are discussed.

\newabstract 

\begin{center}
{\large\bf Lattice QCD Simulations at Small Quark Masses}\\[0.5cm]
{\bf Gerrit Schierholz}\\[0.3cm]
Deutsches Elektronen-Synchrotron DESY\\
D-22603 Hamburg, Germany\\[0.3cm]
John von Neumann-Institut f\"ur Computing NIC\\
Deutsches Elektronen-Synchrotron DESY\\
D-15738 Zeuthen, Germany\\[0.5cm]
-- For the QCDSF Collaboration --
\end{center}

Due to improvement in algorithms and computer performance unquenched
simulations of Wilson-type fermions with lighter quark masses are now
possible~\cite{QCDSF1},\cite{QCDSF2}. This enables us to make contact with chiral
perturbation theory (ChPT) and the real world.

We report results for the pseudoscalar decay constants $f_{PS}^{AB}$ at pion
masses down to 
$O(300)$ MeV, using $N_f=2$ nonperturbatively $O(a)$ improved Wilson fermions.
For calculational details see~\cite{QCDSF3}. The renormalzation constants and
improvement coefficients of the axial vector current are computed
nonperturbatively as well.

We first investigate to see if we are entering a regime, where chiral
logarithms are becoming visible, and look at the ratio of nondegenerate,
partially quenched decay constants~\cite{Sharpe}
\begin{displaymath}
R=\frac{f_{PS}^{VS}}{\sqrt{f_{PS}^{VV} f_{PS}^{SS}}}
\end{displaymath}
($V$:\ valence, $S$:\ sea quark). Our data displays chiral logarithms of about
the expected size. Next we fit our data to the predictions of $NLO$ ChPT. We
obtain $\alpha_4 \approx -0.58$, $\alpha_5 \approx -0.45$. While $\alpha_4$ is
in reasonable agreement with other phenomenological estimates, $\alpha_5$ is
not. Using $r_0=0.5$ fm [$r_0=0.467$ fm] to set the scale, we find
$f_{K^+}/f_{\pi^+} = 1.24$ [$f_{K^+}/f_{\pi^+}=1.22$], to be compared with the
experimental value $1.223$, and $f_{\pi^+}=76(4)(2)$ MeV [$f_{\pi^+}=81(4)(2)$
MeV]. The first error is statistical, while the second error is due to the
error in $r_0/a$.

\newabstract 

\begin{center}
{\large\bf Hadrons for Chirally Improved Fermions}\\[0.5cm]
T. Burch$^1$, C. Gattringer$^2$, L. Y. Glozman$^2$, C. Hagen$^1$, \\ D. Hierl$^1$,
{\bf C. B. Lang}$^2$ and Andreas Sch\"afer$^1$\\[0.2cm]
(BGR [Bern-Graz-Regensburg] Collaboration)\\
$^1$Inst. f. Theoretische Physik, 
Universit\"at Regensburg, D-93040 Regensburg,\\
$^2$Inst. f. Physik, FB Theoretische Physik, Karl-Franzens-Universit\"at Graz, A-8010 Graz,Austria
\\[0.3cm]

\end{center}

We present results for masses of excited mesons \cite{bgr1} and baryons \cite{bgr2} from a quenched calculation with Chirally Improved
quarks at pion masses down to 350 MeV. Our analysis of the correlators is based on the
variational method. It was shown \cite{bgr3} that in this apporach ghost-channels may be identified safely.
The key features are the use of a matrix
of correlators from various source and sink operators and a basis which includes 
Jacobi smeared quark sources with
different spatial widths, thereby improving overlap with states exhibiting radial excitations.
In order to provide a large basis set for spanning the physical states, we also use
interpolators with different Dirac structures. 
Our spectroscopy results for a wide range of ground state and excited hadrons are discussed.

\begin{center}
\includegraphics*[width=0.28\textwidth,clip]{lang1.eps}
\hspace{2mm}
\includegraphics*[width=0.3\textwidth,clip]{lang2.eps}
\hspace{2mm}
\includegraphics*[width=0.3\textwidth,clip]{lang3.eps}
\end{center}
The figures show the results of straightforward extrapolations to the chiral region 
(mesons and positive as well as negative parity baryons).
The horizontal bars represent
the experimental numbers (where known).
Filled symbols are used for those states where no corresponding
state is listed in the particle data summary.

\newabstract 

\begin{center}
{\large\bf Dynamical domain wall fermion simulations and the chiral perturbation theory}\\[0.5cm]
 {\bf Meifeng Lin} \\
Department of Physics, Columbia University,\\
New York, NY 10027, USA\\[0.3cm]
\end{center}

The domain wall fermion formulation has the advantage of possessing
exact flavor symmetry, and chiral symmetry is only mildly broken by a
controllable amount. This allows us to compare the numerical results
from the domain wall fermion simulations directly to the predictions of the
continuum chiral perturbation theory~\cite{Gasser:1984gg}, up to
corrections from the residual chiral symmetry breaking, quantified as
$m_{\rm res}$~\cite{Blum:2000kn}, and possible ${\cal O}(a)$ lattice artifacts. From a
field theoretic point of view, the residual mass $m_{\rm res}$ serves as an
additive mass renormalization to the input quark mass, and the chiral
extrapolations are much simplified compared to the Wilson or Staggered
fermion formulation. For domain wall fermions, the
contamination from higher-dimension operators as a result of the ${\cal
  O}(a)$ lattice artifacts is also exponentially suppressed by the
span of the fifth dimension~\cite{Lin:2006cf}.  And the Symanzik
effective action can be written as
\begin{equation}
  S_{eff} = \int d^4x \lbrack \bar{\psi}(x) ( i\gamma_\mu D^\mu 
- {m_q}) \psi(x) \rbrack + {a e^{-\alpha L_s}} c_{dwf} \bar{\psi}(x)\sigma^{\mu\nu}F_{\mu\nu}\psi(x)
\label{eq:Symanzik}
\end{equation}
where $m_q = m_f + m_{\rm res}$. Guided by this equation, I show that to the next-to-leading order,
including the ${\cal O}(a)$ chiral symmetry breaking term in the
chiral expansions introduces only one new
parameter~\cite{Lin:2006cf}. 

I also present numerical results for the pseudoscalar masses ($M_{PS}$) and decay
constants ($f_{PS}$) from a series of 2+1 flavor dynamical domain wall fermion
simulations carried out by the RBC and UKQCD
collaborations~\cite{Tweedie:2006}. When the Goldstone pion mass is above
400 MeV, the numerical results are not consistent with the NLO chiral
perturbation theory in that simultaneous fits of $M_{PS}^2$ and
$f_{PS}$ fail badly. However, for  pion masses  below 400 MeV,
our preliminary data shows reasonable consistency with the predictions
of NLO ChPT.

\newabstract 

\begin{center}
{\large\bf Applications of Staggered
Chiral Perturbation Theory}\\[0.5cm]
{\bf C.~Aubin}\\[0.3cm]
Dept.\ of Physics, Columbia University,\\
New York, NY USA\\[0.3cm]
\end{center}

One can account for the chiral symmetry violations that arise from staggered fermions on the lattice \cite{SCHPT}. Given current lattice simulations performed by the MILC Collaboration, the $\mathcal{O}(a^2)$ taste-violating errors are of the same order as the light quark masses. Some results using staggered chiral perturbation theory (S$\chi$PT) when applied to light meson physics have been presented in Ref.~\cite{MILCResults}. 

This can be extended to include heavy quark effective theory with S$\chi$PT to understand the chiral behavior of lattice data for heavy-light quarks \cite{HLSCHPT}. From this formalism, several results have been obtained in Ref.~\cite{HLResults}, most notably a prediction of the $D$ decay constant before they were measured by CLEO-c \cite{CLEO}.

Applying these techniques to the kaon mixing parameter $B_K$ is also possible \cite{BKstag}, however it becomes difficult to apply to lattice data due to the large number of free parameters. An option is to used a ``mixed action'' approach, where one uses domain-wall valence quarks and staggered sea quarks \cite{MixedBK}. This technique is a slight modification of the continuum form for $B_K$ and is a promising method with which to extract this quantity.

\newabstract 

\begin{center}
{\large\bf Chiral Perturbation Theory and Lattice QCD with Wilson-type 
Quarks}\\[0.5cm]
{\bf Sinya Aoki}\\[0.3cm]
Graduate School of Pure and Applied Sciences, University of Tsukuba,\\
Ten-oh-dai 1-1-1, Tsukuba, Ibaraki 305-8571, Japan\\[0.3cm]
\end{center}

I explain a necessity of introducing the lattice spacing effect into the
chiral perturbation theory to fit lattice data obtained with Wilson-type
quarks such as pion masses and decay constants. This formulation is
called the Wilson chiral perturbation theory(WChPT)\cite{SA1}. 
The WChPT has been applied to the analysis of the twisted-mass 
QCD \cite{SA2},\cite{SA3},\cite{SA6},\cite{SA8}. 
The twisted-mass QCD becomes automatically
O($a$) improved as long as the twisted-mass is set to the maximal value
(the maximal twist). In actual numerical simulations, however, several 
different definitions of the maximal twist have been employed, and 
different definitions show quiet different chiral behaviours. 
I show that these differences among several definition of the maximal twist
can be understood by the WChPT, and that pion mass and decay constant 
in quenched twisted-mass QCD can be fitted as a function of quark mass
by the WChPT very well.
The WChPT has been extended to $N_f=2+1$ flavor QCD for pseudo-scalar meson
masses\cite{SA4} and for vector meson masses\cite{SA5}. These formula
have been applied to $N_f=2+1$ flavor QCD, in order to fit data. I show
some examples of some chiral fits\cite{SA7}.

\newabstract 

\begin{center}
{\large\bf Lattice QCD with mixed actions:\\
Overlap fermions on a twisted mass sea}\\[0.5cm]
{\bf Oliver B\"ar}$^1$, 
K.~Jansen$^2$, S.~Schaefer$^2$, L.~Scorzato$^3$, A.~Shindler$^2$ \\[0.3cm]
$^1$Institute of Physics, Humboldt University,\\
Newtonstrasse 15, 12489 Berlin, Germany\\[0.3cm]
$^2$NIC, DESY, \\
 Platanenallee 6, 15738 Zeuthen, Germany\\[0.3cm]
 $^{3}$ECT*\\
Strada delle Tabarelle  286, 38050 Villazzano (TN), Italy \\[0.3cm]
\end{center}
We present first results of a mixed action project \cite{Bar:2006zj}.
We analyze gauge configurations generated by the ETM collaboration (for recent summaries see \cite{ShindlerICHEP},\cite{Jansen:2006rf}). Two 
flavors of dynamical twisted mass fermions are taken into account with $m_{\rm PS}\approx300$~MeV. Neuberger's overlap Dirac operator \cite{Neuberger:1997fp} is used for the valence sector. For the quark mass matching we tune the charged valence pion mass such that it equals the charged sea pion mass.

In order to check the partial quenching effects in our mixed action theory we
compute the scalar correlator. As expected, the correlator is negative
\cite{Prelovsek:2004jp}.  Mixed action theories can also be studied with ChPT
methods 
\cite{Bar:2002nr},\cite{Bar:2005tu},\cite{Bar:2003mh}. 
We fit our data to the LO ChPT result for the scalar correlator of Ref.\ \cite{Golterman:2005xa} and find good qualitative agreement.

\newabstract 
\begin{center}
{\large\bf Thoughts on chiral extrapolations for excited hadrons}\\[0.5cm]
{\bf Ulf-G. Mei\ss ner}$^{1,2}$\\[0.3cm]
$^1$HISKP (Th), Universit\"at Bonn, D-53115 Bonn, Germany\\[0.3cm]
$^2$IKP (Th), Forschungszentrum J\"ulich, D-52425 J\"ulich, Germany\\[0.3cm]
\end{center}

An understanding of QCD without ist excitation spectrum is incomplete.
In this talk I review work on chiral extrapolations for excited
hadron states - I concentrate on the $\rho (770)$ and the Roper
$N^* (1440)$. Both these fields are considered as explicit degrees
of freedom in the pertinent chiral effective Lagrangian. The new (large)
scale related to their mass introduces complications in the power counting,
how to treat this in case of the real part of the rho-meson self-energy 
is discussed in \cite{Bruns:2004tj} and the even more complicated
nucleon-Roper-pion system is considered in \cite{Borasoy:2006fk} by
extending the standard infrared regularization scale. The quark mass 
expansion of the $\rho$ looks very similar to the one of the nucleon,
analyzing e.g. the old CP-PACS data, we find $650~{\rm MeV} \leq M_\rho^0
\leq 800~{\rm MeV}$ for the chiral limit mass. For the Roper, the
quark mass expansion comes out very close to the one of the nucleon.
With coupling constants of natural size, no sharp decrease of 
$m_{\rm Roper} (M_\pi)$
for small pion masses is observed. Of course, these considerations need
to be sharpened. For the $\rho$, one has to analyze the more recent data and
in case of the Roper, an extension to include the $\Delta\pi$ and $N (\pi
\pi)_S$ channels is called for. The $\Delta (1232)$ resonance is discussed
in  Bernard's talk \cite{VB} and for dynamically resonances I refer 
to Oller's talk \cite{JAO}.

\newabstract 

\begin{center}
{\large\bf Chiral extrapolation of baryon properties}\\[0.5cm]
{\bf V. Bernard} \\[0.3cm]
Laboratoire de Physique Th\'eorique, 67085 Strasbourg, France\\[0.3cm]
\end{center}

\noindent
Enormous progresses have been made in lattice calculation in the last
years. However these calculations still involve rather large pion masses. 
It is thus necessary to perform chiral extrapolations in order to make contact
to the physical world. This can be done in a model independent manner using
chiral perturbation theory. For discussion on the different regularization
used in baryon CHPT see \cite{ta}. 
In this talk I reported on chiral extrapolation 
of some baryon properties, namely the mass \cite{mcgov} and  
axial vector coupling \cite{bem} of the nucleon
and the $\Delta$ mass \cite{bhm} essentially in the continuum limit. 
In the nucleon case two
loop calculations are
by now available (only the $O(p^5)$ terms have been determined in the case of the mass).
In these calculations only 7 combinations of LEC's from the $\pi N$ Lagrangian 
up to dimension 4 appear,
5 of these
being rather well determined from $\pi N$ scattering and the Goldberger
Treiman discrepancy. Taking into account the error bars on these LECs and 
studying the convergence of the series one finds that the chiral extrapolation
of these two quantities can be trusted up to $\sim 350$ MeV. Taking explicitely
into account the $\Delta$ degree of freedom does not improve on the result
\cite{phw}. Discussion on  finite volume effects can be found for exemple in \cite{ch}.
In the case of the $\Delta$ mass  fits to the lattice data in the small
scale expansion  seems to 
indicate that the symmetry breaker LEC $a_1$ is far from its SU(6) value
leading to a $\pi \Delta$ sigma term smaller than the $\pi N$ one. Further
study is needed to confirm this result. The conclusion of these studies 
is that it is not usefull to fit lattice data up to rather large pion
masses but that it is much more important to  perform {\it simultaneous systematic 
chiral extrapolation with a consistent set of LEC's}
and  to carefully {\it evaluate the theoretical uncertainties}.

\newabstract 
\begin{center}
{\large\bf Some topics in manifestly Lorentz-invariant ChPT}\\[0.5cm]
D.~Djukanovic$^1$,
J.~Gegelia$^1$, {\bf M.~R.~Schindler}$^1$ and S.~Scherer$^1$\\[0.3cm]
$^1$Institut f\"ur Kernphysik, Johannes Gutenberg-Universit\"at\\
55099 Mainz, Germany\\[0.3cm]
\end{center}

   We present three topics in manifestly Lorentz-invariant baryon
chiral perturbation theory (BChPT).

   The higher-derivative formulation of ChPT
\cite{Djukanovic:2004px} presents an alternative regularization
scheme that can be used in the vacuum, one- and few nucleon sector
of ChPT.
   By performing field transformations in the canonical
Lagrangian, additional higher-derivative terms are introduced in
the lowest-order Lagrangian.
   These terms improve the ultraviolet behavior of propagators
while preserving all symmetries.
   The new parameters can be interpreted as smooth cutoffs,
whose values can be chosen freely.

   To investigate the applicability of chiral expansions \cite{Djukanovic:2006xc} we
compare the sum of an infinite number of terms contributing to the
nucleon self energy starting at the two-loop level with the
leading non-analytic term $\sim M^3$.
   For pion masses $M>550\,\mbox{MeV}$ the contributions from the sum
of higher-order terms is larger than the leading-order
contribution.
   This shows that for $M>550\,\mbox{MeV}$ the power counting is no
longer valid.

   We also present the results of the first complete two-loop
calculation of the nucleon mass \cite{SchindlerPrep} using the
reformulated infrared renormalization \cite{Schindler:2003xv} and
compare our result to the heavy baryon result at order ${\cal
O}(q^5)$ \cite{McGovern:1998tm}.
   While at order ${\cal O}(q^5)$ only low energy constants appear
that have previously been determined, the terms at order ${\cal
O}(q^6)$ contain combinations of numerically unknown constants.

\newabstract 

\begin{center}
{\large\bf Chiral Extrapolations in Few--Nucleon Systems}\\[0.5cm]
{\bf Evgeny Epelbaum}$^{1,2}$\\[0.3cm]
$^1$Institut f\"ur Kernphysik, Forschungszentrum J\"ulich, D-52425 J\"ulich, Germany\\[0.1cm]
$^2$HISKP (Theorie), Univertit\"at Bonn, Nu\ss{}allee 14-16, 53115 Bonn, 
Germany
\end{center}

The dependence of the low-energy dynamics of QCD on the variation of the 
up and down quark masses can be naturally studied within the framework of
chiral effective field theory. While the situation in the Goldstone boson and
single-nucleon sectors is rather straightforward due to the validity of
perturbation theory, it is highly non-trivial in the few-nucleon sector where
non-perturbative methods are required to  understand e.g.~the properties of the
shallow bound states. The quark mass dependence of various few-nucleon observables
is not only of academic interest, but also relevant for interpolating the results from
lattice gauge theory \cite{Fukugita:1994ve}, \cite{Beane:2006mx} and imposing bounds on the time-dependence
of fundamental couplings.   

We have studied the quark mass dependence of the nucleon-nucleon (NN) force
and various two-nucleon observables at
next-to-leading order in the chiral expansion \cite{Epelbaum:2002gb}, see also \cite{Beane:2002vs}
for a similar work. We found that both NN S-wave scattering lengths become
smaller in magnitude (i.e.~more natural) in the chiral limit. The deuteron
is less bound when the quark mass is increased and becomes unbound for $M_\pi$ 
larger than $\sim 200$ MeV. The theoretical uncertainty of our results is
rather large and mainly caused by the uncertainty
in the determination of the LEC $\bar d_{16}$ and by the unknown LECs
associated with the leading $M_\pi$-dependent NN contact interactions.  
We have also studied the infrared renormalization group limit cycle in the 
three-nucleon system \cite{Epelbaum:2006jc}, whose existence for certain values of $m_u$ and $m_d$ 
was conjectured in \cite{Braaten:2003eu}.

\newabstract 

\begin{center}
{\large\bf Nuclear Lattice Simulations}\\[0.5cm]
{\bf B. Borasoy}\\[0.3cm]
Helmholtz-Institut f\"ur Strahlen- und Kernphysik (Theorie) \\
Universit\"at Bonn, Nu{\ss}allee 14-16, D-53115 Bonn, Germany \\[0.3cm]
\end{center}

Nuclear lattice simulations are a novel approach to nuclear physics based
on numerical simulations of chiral effective field theory on the lattice.
Within this scheme the chiral effective Lagrangian is formulated on the
lattice and the corresponding path integral is evaluated numerically by
Monte Carlo sampling.

In two exploratory studies we have employed this framework to calculate
properties of the lightest nuclei: the deuteron, triton
and Helium-4. In \cite{BKLM} the triton was investigated in the SU(4) limit 
of the pionless theory including the three-nucleon force. By comparing with
continuum  results we demonstrated that the nuclear lattice
formalism can be used to study few-body nucleon physics. A first
measurement of the three-nucleon force on the lattice in the Wigner
symmetry limit has been provided. Moreover, we have shown that many-body simulations of cold
dilute nuclear matter in the Wigner symmetry limit should be possible
without a sign problem.
This work is thus also of relevance for
future many-body simulations with arbitrary numbers of nucleons including
three-body effects.

The pionful theory is considered in
\cite{BEKLM} where we calulate the binding energies
and sizes of triton and $^4$He. A Monte Carlo algorithm with pions and
auxiliary fields is constructed which reproduces both the lowest order
$S$-wave contact interactions and instantaneous one-pion exchange between
the nucleons. Due to the approximate Wigner symmetry of the actual
interactions we find only minor sign and phase oscillations for light
nuclei. The importance of higher order interactions is also investigated.

Although our investigations are still at the exploratory level, lattice
simulations of chiral effective field theory appear to be a promising tool
to investigate few- and many-body nuclear physics with a clear theoretical
connection to QCD and a systematic chiral expansion.

\newabstract 
\begin{center}
{\large\bf Chiral Non-Perturbative Study of Pseudoscalar Self-Energies}\\[0.5cm]
Luis Roca and {\bf Jos\'e Antonio Oller}\\[0.3cm]
Departamento de F\'{\i}sica, Universidad de Murcia,\\
E-30071 Murcia, Espa\~na\\[0.3cm]
\end{center}

In this talk we report about our recent work \cite{masses}. 
We perform a non-perturbative chiral study of the masses of the lightest
pseudoscalar mesons.  The pseudoscalar self-energies
are calculated by the evaluation of the scalar self-energy loops with
full S-wave meson-meson amplitudes taken from Unitary Chiral Perturbation Theory
(UCHPT) \cite{nd}. These amplitudes, among other features, 
contain the lightest nonet of scalar resonances $\sigma$,
$f_0(980)$, $a_0(980)$ and  $\kappa$ and the heavier ones up to 1.5 GeV.
 The self-energy loops are regularized
by a proper subtraction of the infinities within a dispersion
relation formulation of the scattering amplitudes. Values for the
bare masses of pions and kaons are obtained as well as  an estimate
of the mass of the $\eta_8$. We then match to the self-energies
from standard Chiral Perturbation Theory (CHPT) to  $\Opc$ and
resum higher orders from our calculated scalar self-energies. The
dependence of the self-energies on the quark masses allows  a
determination of the ratio of the strange quark mass upon the
mean of the lightest quark masses,  $m_s/\hat{m}$,  in terms of 
the $\Opc$ CHPT low energy constant combinations $2L^r_8-L^r_5$ and
$2L^r_6-L^r_4$.  In this way, we give a range for the values of these
low energy counterterms and for $3L_7+L^r_8$,  once the $\eta$ meson mass
is invoked. The low energy constants are further constraint by
performing a fit to the recent MILC lattice data on the pseudoscalar
masses. An excellent reproduction  of the MILC data is obtained, 
at the level of  $1$\% of relative error in the pseudoscalar masses, 
 and  $m_s/\hat{m}=25.6\pm 2.5$ results. This value
is consistent with $24.4\pm 1.5$ from CHPT and phenomenology 
  and more marginally with the value $27.4\pm
0.5$ obtained from pure perturbative chiral  extrapolations of the MILC
lattice data to physical values of the lightest quark masses.   We also
show that our estimation of the ${\cal O}(p^6)$ and higher chiral 
orders, indicates that the SU(3) CHPT series on pseudoscalar
self-energies seems to stabilize and behave much better once 
the ${\cal O}(p^6)$ CHPT contributions to the pseudoscalar 
masses are included, despite that the ${\cal O}(p^4)$ contributions are 
typically smaller than the ${\cal O}(p^6)$ ones.

\newabstract 

\begin{center}
{\large\bf Scalar two-loop diagrams on the lattice}\\[0.5cm]
 {\bf Hermann Krebs}\\[0.3cm]
HISKP, Universit\"at Bonn
and
IKP, Forschungszentrum J\"ulich\\[0.3cm]
\end{center}

Numerical simulations on the lattice are always performed in finite volume
with a finite lattice spacing. In order to understand both the infinite volume 
and the continuum limit non-trivial extrapolations are needed. 
In some cases lattice
perturbation theory provides a powerful tool in 
 getting an analytical control over nonperturbative
simulations. 

Lattice perturbation theory itself is a challenging field. Since 
Lorentz-invariance is lost on the lattice many tools developed for continuum
 perturbation theory such as Feynman parameters are not applicable on the
 lattice. 
Especially for multi-loop calculations where conventional methods often lead 
to loss in precision new techniques are highly desirable. A novel powerful 
technique in this respect was introduced by L\"uscher and Weisz~\cite{LW}. 
Instead of integration over the first Brillouin-zone in momentum space they 
suggested to calculate infinite sums in coordinate space. In this case 
extensive knowledge of the free lattice propagators is required. 
Recursion relations for the massless propagator were derived which 
allowed to express the Green function at any lattice site as a linear 
combination of two constants which can be determined to very high precision. 

We extended this technique to the massive case and could give the small mass 
expansion of the massive scalar propagator~\cite{BK1}. In the BPHZ subtraction 
scheme any scalar two loop diagram can be expressed as a sum of a regular and 
a singular part which converge and diverge in the continuum limit,
respectively. The singular part is always of the following sunset type:
\begin{equation}
\int_{-\pi}^{\pi}\frac{d^4k}{(2\pi)^4}\frac{d^4q}{(2\pi)^4}
\frac{\prod_\mu\widehat{k}_\mu^{2i_\mu}\cdot\widehat{q}_\mu^{2j_\mu}\cdot
\widehat{(k+q)}_\mu^{2l_\mu}}{[\widehat{k}^2+m^2]^\alpha
[\widehat{q}^2+m^2]^\beta[\widehat{(k+q)}^2+m^2]^\gamma},
\quad \hat{k}_\mu=2 \sin\left(\frac{k_\mu}{2}\right).
\end{equation}
We developed a method for deriving the small mass expansion of any singular part~\cite{BK2}.
The coefficients of this expansion can be calculated numerically to very high
precision such that even discretization effects can also be studied within our method.
The proposed method for the small mass expansion is universal and can 
be applied once the particle propagator is known, e.g., 
it can easily be extended to Wilson fermions.

\newabstract 
\begin{center}
{\large\bf Nucleon and N to {\boldmath$\Delta$}  form factors in Lattice QCD}\\[0.5cm]
 {\bf C. Alexandrou}$^1$, Th. Leontiou$^1$, 
J. W. Negele$^2$ and A. Tsapalis$^3$\\[0.2cm]
$^1$Dep. of Physics, University of Cyprus, CY-1678 Nicosia, Cyprus\\[0.2cm]
$^2$ Dep. of Physics, M.I.T, Cambridge, Massachusetts 02139, U.S.A.,\\[0.2cm]
$^3$I.A.S.A., University of Athens, Athens, Greece.
\end{center}

The isovector 
nucleon form factors are evaluated in lattice QCD in the quenched theory 
and  using two degenerate flavors of dynamical
Wilson fermions for  a range of pion masses between 690 and 380 MeV,
 using the nucleon mass to set the lattice spacing $a$. We 
 find that the  isovector electric form factor,
 $G_E$, at the physical limit  deviates more from experiment 
than the magnetic form factor, $G_M$~\cite{NN}. Lattice results show a
weaker $q^2$-dependence for 
the isovector ratio $\mu G_E/G_M$   as compared to the
results obtained in recent polarization experiments~\cite{expNN}.
The electromagnetic and axial  N to $\Delta$ transition form factors
are evaluated for quenched and two dynamical flavor of Wilson fermions as
well as in a hybrid scheme where we use MILC configurations and domain
wall fermions over a range of pion masses between 690 and 360 MeV.
We find that lattice results  for  the dominant magnetic dipole form factor
 in $\gamma^* N \rightarrow \Delta$, when linearly extrapolated in $m_\pi^2$
to the
 physical point, yield larger values than experiment~\cite{NDelta}.
Results for the 
ratios of electric and coulomb quadrupole amplitudes  to the
magnetic dipole, EMR and CMR, are non-zero and negative as in
 experiment albeit with statistical errors that
are still too large in
the unquenched case~\cite{NDelta2} to allow assessment of pion cloud 
contributions.
Our results for the four N to $\Delta$ axial form factors 
show that $C^A_3\sim 0$, $C^A_4$ is small and
 $C^A_5$ and $C^A_6$  are the dominant form factors. 
We evaluate
 the ratio $C^A_5/C^V_3$, which is the analog of $g_A/g_V$,
  as a function
of $Q^2$. This ratio provides  the leading contribution to the
 the parity violating asymmetry~\cite{NDaxial} to be measured by the G0 
experiment at JLab~\cite{G0}. The off-diagonal Goldberger-Treiman relation
is examined by evaluating, at the same lattice parameters, $g_{\pi N\Delta}$ and
$f_\pi$.

\newabstract 

\begin{center}
{\large\bf Hadron Structure using DWF Quarks on an Asqtad Sea}\\[0.5cm]
{\bf David Richards}, for Lattice Hadron Physics Collaboration\\[0.3cm]
Jefferson Laboratory, 12000 Jefferson Avenue, Newport News, 
VA 23606, USA.\\[0.3cm]
\end{center}

Moments of unpolarized, helicity, and transversity distributions,
electromagnetic form factors, and generalized form factors of the
nucleon are presented from a preliminary analysis of lattice results
using pion masses down to 359 MeV\cite{Edwards:2006qx}.  We employ a
hybrid approach, in which improved, staggered quarks are used for the
generation of the gauge configurations, whilst domain-wall fermions,
with their desirable chiral properties, are used for the valence
quarks.

The nucleon axial-vector charge, a benchmark quantity of QCD, is
particularly robust under chiral extrapolation; the consistency of the
hybrid calculation, both with other lattice calculations, and with
experiment at the physical pion mass, is
encouraging\cite{Edwards:2005ym}.  Lattice moments of
structure functions and GPDs likewise require extrapolation to the
physical quark masses; a long-standing puzzle has been the flat
behaviour of the flavour-non-singlet momentum fraction, $\langle x
\rangle$, of the nucleon, at a value considerably higher than the
experimental value.  An approach in which we apply $\chi$PT, with
low-energy constants $g_A$ and $f_\pi$ given by their lattice values
at each quark mass, allows a two-parameter extrapolation in
$m_\pi^{\rm lat}/f_\pi^{\rm lat}$ to yield a value for $\langle x
\rangle$, and other benchmark quantities, at the physical quark masses
that are consistent with experiment.  This development encourages to
now exploit the predictive power of these calculations.

The low-$Q^2$ behaviour of the nucleon form factors describes the
distribution of charge and magnetism within a nucleon. The slope of
the $F_1$ form factor is related to the rms charge radius; the chiral
extrapolation of the isovector charge radius likewise yields values
consistent with experiment\cite{Dunne:2001ip}.  Generalized Parton
Distributions provide new insight to hadron structure.
For example, the total angular momentum carried by the quarks is related
to a combination of moments $J_q = \frac{1}{2}(A^{u+d}_{20} +
B^{u+d}_{20})$\cite{Ji:1996ek}. Combined with measurements of
quark spins, we find the total orbital angular
momentum carried by quarks is small, though that carried by individual
flavours is substantial.

\newabstract 
\begin{center}
{\large\bf Distribution amplitudes from the lattice}\\[0.5cm]
V.~M.~Braun$^1$, M.~G{\"o}ckeler$^1$, R.~Horsley$^2$,
H.~Perlt$^3$, D.~Pleiter$^4$, P.~E.~L.~Rakow$^5$,
  G.~Schierholz$^{36}$, A.~Schiller$^3$,
  \textbf{W.~Schroers}$\,^4$,
  H.~St{\"u}ben$^7$, and J.~M.~Zanotti$^2$\\[0.3cm]
QCDSF/UKQCD collaboration] \\[0.3cm]
$^1$Institut f\"ur Theoretische Physik, Universit\"at
Regensburg, \\93040 Regensburg, Germany\\[0.3cm]
$^2$School of Physics, University of Edinburgh,\\
  Edinburgh EH9 3JZ, UK\\[0.3cm]
$^3$Institut f{\"u}r Theoretische Physik,
  Universit{\"a}t Leipzig, \\04109 Leipzig, Germany\\[0.3cm]
$^4$John von Neumann-Institut f\"ur Computing NIC /
  DESY, \\15738  Zeuthen, Germany\\[0.3cm]
$^5$Theoretical Physics Division, Department of Mathematical
  Sciences, University of Liverpool,\\ Liverpool L69 3BX, UK\\[0.3cm]
$^6$Deutsches Elektronen-Synchrotron DESY,\\
  22603 Hamburg, Germany\\[0.3cm]
$^7$Konrad-Zuse-Zentrum f\"ur Informationstechnik Berlin, \\14195
  Berlin, Germany[0.3cm]
\end{center}

\newcommand{\msbar}{$\overline{\mbox{\rm MS}}$}  
Hadronic wave functions are of crucial importance when describing
exclusive and semi-ex\-clu\-si\-ve reactions~\cite{Brodsky:1989pv}.
For detailed references and applications consult~\cite{Braun:2006dg}.
Distribution amplitudes (DAs) are related to the hadron's
Bethe-Salpeter wave function, $\phi(x,k_\perp)$, by an integral over
transverse momenta.  For the leading twist meson-DAs we have
\begin{equation}
  \label{eq:dadef}
  \phi(x,\mu^2) = Z_2(\mu^2)\int_{\vert k_\perp\vert<\mu} d^2k\,
  \phi(x,k_\perp)\,,
\end{equation}
where $x$ is the quark longitudinal momentum fraction, $Z_2$ the
renormalization factor (in the light-cone gauge) for the quark-field
operators in the wave-function, and $\mu$ denotes the renormalization
scale. In this presentation we quote all numbers with a scale
$\mu^2=4\,$ GeV$^2$ in the \msbar-scheme.

It is convenient to rescale $\xi=2x-1$. It is common to expand DAs
into their Gegenbauer moments and quote the expansion coefficients,
$a_i$, at a given renormalization scale as a parameterization of DAs,
\begin{equation}
  \label{eq:daexpansion}
  \phi(\xi,\mu^2) = \frac{3}{4}(1-\xi^2)\left( 1+\sum_{n=1}^\infty
  a_n(\mu^2) C_n^{3/2}(\xi) \right)\,.
\end{equation}
The zeroth moment is normalized to unity, $\int_{-1}^1d\xi
\phi(\xi,\mu^2)=1$, at any energy scale $\mu^2$. Taking the $u$- and
$d$-quarks to be degenerate, $G$-parity implies that the pion DA is an
even function of $\xi$ and hence all odd moments vanish, i.e.,
$a^\pi_{2n+1}=0$.

Recently, we have computed the first moments of meson distribution
amplitudes~\cite{Braun:2006dg} on the lattice. Independently, a
calculation of the first moment of the kaon distribution amplitude has
appeared which uses a different discretization scheme and different
working points~\cite{Boyle:2006pw}. We find that their results are
compatible with ours.

This presentation discusses the calculation of the first non-vanishing
moment of the pion DA, $a^\pi_2$, and the first two moments of the
kaon DA, $a^K_1$ and $a^K_2$. We compare our results to previous
estimates from sum rules and experiment and discuss the
phenomenological implications of our lattice computation.

We obtain $a_2^\pi(4\,\mbox{GeV}^2) = 0.201(114)$ which disfavors the
commonly adopted ``asymptotic'' model with $a_{2n}^\pi=0$. This result
is also compatible with model estimates,
cf.~e.g.~\cite{Bakulev:2004ar}.

We also find $a_2^K(4\,\mbox{GeV}^2) = 0.175(18)(47)$ which provides
an important test of different and competing kaon models employed in
the literature, see~\cite{Ball:2006wn}.

Finally, we find $a_1^K(4\,\mbox{GeV}^2) = 0.0453(9)(29)$ which not
only confirms the sign, but also the order of magnitude for the
asymmetry of the kaon wave function. This key finding is of major
importance both for the understanding of $B$ physics and of weak
decays~\cite{Ball:2006wn}.

\newabstract 

\begin{center}
{\large\bf Pion structure from lattice QCD}\\[0.5cm]
{\bf{D.~Br\"ommel}}$^{1,2}$,  M.~Diehl$^1$, M.~G\"ockeler$^2$, Ph.~H\"agler$^3$,\\
R.~Horsley$^4$, Y.~Nakamura$^5$, D.~Pleiter$^5$, P.E.L.~Rakow$^6$,\\
A.~Sch\"afer$^2$, G.~Schierholz$^{1,5}$, H.~St\"uben$^7$ and J.M.~Zanotti$^4$\\[0.3cm]
$^1$Deutsches Elektronen-Synchrotron DESY, Hamburg, 
$^2$Universit\"at Regensburg, 
$^3$Technische Universit\"at M\"unchen, 
$^4$University of Edinburgh, 
$^5$John von Neumann-Institut f\"ur Computing NIC/ DESY, Zeuthen, 
$^6$University of Liverpool, 
$^7$Konrad-Zuse-Institut f\"ur Informatik ZIB, Berlin\\[0.3cm]
\end{center}

We calculate moments of generalised parton distributions (GPDs) of the pion.
This is done using a simulation of non-perturbatively improved dynamical Wilson
fermions with two flavours with Wilson glue. The configurations have been
generated within the QCDSF/UKQCD/DIK collaborations and cover a range of pion
masses down to $\sim\!340\,\textrm{Me}\kern-0.06667em\textrm{V\/}$.
We show partly preliminary results for the lowest (two) moment(s) of the tensor
(vector) GPD $H^q_{T,\pi}$ ($H^q_{\pi}$), defined as
\begin{eqnarray}
\label{vgpd}\frac{2 P^{\mu}}{n\cdot P} H^q_{\pi} =
      \int\frac{\mathrm{d}\lambda}{2\pi}
      \mathrm{e}^{\mathrm{i} \lambda n\cdot P x}
      \langle \pi(p') | \overline{q}(-\frac{\lambda}{2}n)\,
      \gamma^{\mu}\,\mathcal{U}\, q(\frac{\lambda}{2}n) | \pi(p)\rangle +
      \mathrm{higher~twist}\,,\\
\label{tgpd} \frac{P^{[\mu}\Delta^{\nu]}}{m_{\pi}\,n\cdot P}H^q_{\mathrm{T},\pi}
      = \int\frac{\mathrm{d}{\lambda}}{2\pi}
      \mathrm{e}^{\mathrm{i} \lambda n\cdot P x}
      \langle \pi(p') | \overline{q}(-\frac{\lambda}{2}n)\,\mathrm{i}
      \sigma^{\mu\nu}\,\mathcal{U}\, q(\frac{\lambda}{2}n) | \pi(p) \rangle +
      \mathrm{higher~twist}\,. 
\end{eqnarray}
The moments of these GPDs are parametrised by generalised form factors. We focus
on the lowest moment of the vector GPD Eq.~(\ref{vgpd}) which is the pion form
factor $F_{\pi}$.  Details of the calculation and the lattice setup and further
references can be found in \cite{us1},\cite{us2}.
The pion form factor can be very well described by a monopole ansatz. The
squared monopole masses obtained from this ansatz are extrapolated to the
physical point linearly against square of the pion mass. From that we obtain a
charge radius of $\langle r^2 \rangle = 0.440(19)$~fm$^2$ in good agreement with
experiment. We also perform a study of the finite size and scaling effects
\cite{us2}.
The second moment of the vector GPD corresponds to the momentum fraction of the
quarks inside the pion. A preliminary result indicates that quarks and
anti-quarks inside the pion carry $\sim\!50$\% of the momentum fraction.
The calculation of the pion tensor GPD Eq.~(\ref{tgpd}) has not been attempted
before. This quantity is experimentally unknown and thus provides genuine new
insights to the internal spin structure of the pion \cite{tvs}. Our preliminary
data so far indicate a rich structure with a non-vanishing tensor GPD also in
the forward limit.%

\newabstract 

\begin{center}
{\large\bf Nucleon Structure Functions and Form Factors}\\[0.5cm]
{\bf Dirk Pleiter}$^1$\\[0.3cm]
For the QCDSF collaboration\\[0.3cm]
$^1$Deutsches Elektronen-Synchrotron DESY, 15738 Zeuthen, Germany\\[0.3cm]
\end{center}

In this talk we have reported on recent results for the moments of the
unpolarised nucleon structure functions \cite{Gockeler:2004vx}
and the electromagnetic form factors \cite{lat06}.
The calculations have been done using dynamical gauge field
configurations with $N_{\rm f}=2$ flavours of non-perturbatively
$O(a)$-improved Wilson fermions, which have been generated
by the QCDSF, DIK and UKQCD collaborations.
On the different gauge field ensembles the pseudo-scalar meson mass is
in the range of 340 to 1170 MeV. The lattice spacings vary between 0.07
and 0.11 fm, i.e.~our lattice are relatively fine. The spatial extension
of our lattice is 1.5 fm for the heavy sea quark masses and up to 2.6 fm
for the very light quark mass. The renormalisation constants have been
determined non-perturbatively.

By comparing our results for different lattice spacings and different
volumes we find that discretisation and finite volume effects are relatively
small. The systematic error which seems most difficult to control stems
from the extrapolation of the lattice results to the physical
mass of the up/down quarks.  When using a naive ansatz linear in
the quark mass, which describes our data actually very well, we
find a significant discrepancy with the experimental values. Using
the quenched approximation we previously obtained surprisingly
similar results \cite{Gockeler:2004wp},\cite{Gockeler:2003ay}.

For the moments of the structure functions as well as for the form
factor radii and anomalous magnetic moments chiral effective theories
have been used to investigate the quark mass dependence. These calculations
suggest a rather strong quark mass dependence at very small quark masses.
The relevant range of quark masses is only starting to become accessible
for lattice simulations.

\newabstract 

\begin{center}
{\large\bf {\boldmath$B_K$} from dynamical Domain Wall Fermion Simulations}\\[0.5cm]
{\bf Enno E. Scholz}\\for the RBC- and UKQCD-collaborations\\[0.3cm]
Brookhaven National Laboratory\\
Upton, NY 11973, USA
\end{center}

The kaon bag-parameter, 
%
$B_K\;=\;\langle K^0|\bar s\gamma_\mu(1-\gamma_5)d\bar 
s\gamma_\mu(1-\gamma_5)d|\bar K^0\rangle/(\frac83f_K^2m_K^2),$
defined as the ratio of the $\Delta S=2$ matrix element to its value obtained
in 
the vacuum saturation approximation 
is a measure of indirect $\mathsf{CP}$-violation and provides an important
 non-perturbative input to the Cabbibo-Kobayashi-Maskawa matrix fit,
 cf.\ e.g.\ \cite{Bona:2005vz}. 

We present a preliminary study of the kaon bag-parameter $B_K$ measured on 
$L^3 \times T \times L_s=16^3\times32\times16$ lattices using $N_f=2+1$
dynamical 
flavors of domain wall fermions, see also 
\cite{Cohen:2006xi}, \cite{SaulLattice06}, \cite{RBCUKQCD:inPrep}. The mass of the heavier 
(single) flavor was tuned to match the physical strange quark mass, while
three 
different values for the masses of the two degenerate light quarks have 
been used in the simulation. These correspond to pion masses in the range
 of 390 to 630 MeV. Whereas the applicability of chiral perturbation theory 
to extrapolate the pion masses and decay constants to the physical limit is at 
least questionable in this mass range (see \cite{Lin:2006cf},
\cite{MeifengTrentoProc} 
for a discussion), we still may use the partially quenched chiral perturbation 
theory formulae given in \cite{VandeWater:2005uq} as a smooth fitting function 
for an extrapolation of our results concerning $B_K$. Ignoring small 
non-degenerate effects even allows for an interpolation over just a small
 mass range. For details we refer to \cite{SaulLattice06} and 
our upcoming paper \cite{RBCUKQCD:inPrep}.

\newabstract 
\begin{center}
{\large\bf Low-energy constants for {\boldmath$\Delta{S}=1$} transitions}\\[0.5cm]
L.\,Giusti$^1$, P.\,Hern\'andez$^2$, M.\,Laine$^3$, C.\,Pena$^1$,
P.\,Weisz$^4$,
J.\,Wennekers$^5$ and {\bf H.\,Wittig}$^6$\\[0.2cm]
$^1$CERN, Department of Physics, TH Division, CH-1211 Geneva 23\\[0.2cm]
$^2$Dpto. F\'{\i}sica Te\'orica and IFIC, E-46071 Valencia\\[0.2cm]
$^3$Faculty of Physics, University of Bielefeld, D-33501
    Bielefeld\\[0.2cm]
$^4$Max-Planck-Institut f\"ur Physik, F\"ohringer Ring 6, D-80805
    Munich\\[0.2cm] 
$^5$DESY, Notkestra{\ss}e 85, D-22603 Hamburg\\[0.2cm]
$^6$Institut f\"ur Kernphysik, Universit\"at Mainz, D-55099 Mainz
\end{center}

In order to investigate the origins of the $\Delta{I}=1/2$ rule in
$K\to\pi\pi$ decays we have devised a strategy to determine the LECs
of operators describing $\Delta{S}=1$ transitions \cite{strat}. We
keep an active charm quark, express $K\to\pi\pi$ amplitudes in terms
of LECs and study their dependence on $m_c$. By employing Neuberger
fermions, the severe problem of mixing with lower dimensional
operators is completely avoided. The LECs are extracted by matching
lattice QCD data for suitable ratios of correlation functions to the
expressions of ChPT in finite volume\,\cite{HerLai06}. The first step
in our implementation of the strategy is focused on the case of a
light and degenerate charm quark, i.e. the GIM limit ($m_c=m_u$). In
order to tame the extreme statistical fluctuations encountered in
simulations near the chiral limit, noise reduction techniques
(low-mode averaging) must be employed\,\cite{jljl}. After
renormalising ratios of correlators non-perturbatively \cite{NPren},
we perform joint fits to the corresponding ChPT expressions in both
the $\epsilon$ and $p$-regimes. Our results in the GIM limit
\cite{GIM} indicate a previously unseen, significant enhancement of
the $\Delta{I}=1/2$ amplitude, which however, is still a factor~4
smaller than the experimental number. First attempts to analyse the
direct influence of the charm quark in ChPT \cite{HerLai04} will be
supplemented by future numerical studies.

\newabstract 
\begin{center}
{\large\bf The \boldmath{$\Delta$}-resonance in a finite volume}\\[0.5cm]
Veronique Bernard$^1$, Ulf-G. Mei\ss ner$^{2,3}$ and {\bf Akaki Rusetsky}$^{2,4}$\\[0.3cm]
$^1$Universit\'e Louis Pasteur, Laboratoire de Physique Th\'eorique\\
3-5, rue de l'Universit\'e, F-67084 Strasbourg, France\\[0.3cm]
$^2$Universit\"at Bonn, Helmholtz-Institut f\"ur Strahlen- und Kernphysik \\
Nu\ss allee 14-16, D-53115 Bonn, Germany\\[0.3cm]
$^3$Forschungszentrum J\"ulich, Institut f\"ur Kernphysik (Theorie)\\
D-52425 J\"ulich, Germany\\[0.3cm]
$^4$High Energy Physics Institute, Tbilisi State University\\
University st. 9, 380086 Tbilisi, Georgia\\[3mm]
\end{center}

The aim of the investigation, carried out in the present work, is 
to study the feasibility of the extraction of the $\Delta$-resonance
parameters from the lattice data at the physical value of the
pion mass, i.e. in the case of unstable $\Delta$. To this end, we calculate
the self-energy of the $\Delta$-resonance in a finite volume
up to and including $O(p^3)$ in Chiral Perturbation Theory, using infrared 
regularization. The
poles of the $\Delta$-propagator, obtained in these calculations,
determine the energy spectrum of the correlator of two $\Delta$-fields
in a finite Euclidian box and can be obtained in lattice QCD.
The results of our calculations enable one 
to investigate the dependence of the energy levels
both on the size $L$ of the box and on the pion mass $M_\pi$.  

As previously argued in the literature 
(see, e.g.~\cite{Luescher},\cite{Wiese},\cite{DeGrand}),
the presence of a narrow resonance in the particle spectrum
manifests itself through a peculiar irregular
behavior of the energy levels with respect to $L$ (avoided level crossing
near the resonance energy). We have however 
verified by explicit calculations at 
$O(p^3)$ that, in the case of interest, the dependence of energy levels is
rather smooth -- the avoided level crossing 
is nearly washed out due to a large 
$\Delta$-width. Despite this fact, on the basis of the numerical study,
we argue that the extraction of the 
parameters of $\Delta$ (namely, of the mass and the $\pi N\Delta$ coupling 
constant) from a measured $L$-dependence of the lowest energy levels is 
still a feasible task.

\newabstract 

\begin{center}
{\large\bf Lattice QCD on Smaller and Larger Volumes}\\[0.5cm]
{\bf Meinulf G\"ockeler}$^1$\\[0.3cm]
For the QCDSF collaboration \\[0.3cm]
$^1$Institute for Theoretical Physics, University of Regensburg \\
93040 Regensburg, Germany.\\[0.3cm]
\end{center}

The QCDSF collaboration (together with the DESY-ITEP-Kanazawa (DIK)
collaboration) is generating gauge field configurations with
two degenerate flavours of nonperturbatively $O(a)$-improved Wilson
fermions (clover fermions) and Wilson glue, including in the analysis
also older ensembles from UKQCD. In this talk we have 
discussed our approach to the treatment of finite size effects.
It is based on appropriate versions of chiral effective field theory
and hence intimately connected with the quark mass dependence of the
physical quantities considered. In particular we deal with data 
from our two finite size runs (three volumes each, with the other 
simulation parameters kept fixed). 
Ongoing simulations at smaller quark masses are expected to shed new
light on the chiral behaviour (see, e.g., Refs.~\cite{roger},\cite{yoshi}).
 
Among the quantities studied are the nucleon mass~\cite{finitemass} and 
the axial coupling constant of the nucleon~\cite{gapap}. For the 
pion (or pseudoscalar) mass and the pion decay constant we make use 
of the recently proposed ``resummed'' L\"uscher type 
formulae~\cite{cola}.

\newabstract 

\begin{center}
{\large\bf Utilizing covariant BChPT for chiral Extrapolations}\\[0.5cm]
{\bf Tobias A. Gail} and Thomas R. Hemmert\\[0.3cm]
Physik-Department, Theoretische Physik T39\\
        TU-M\"unchen, D-85747 Garching, Germany\\[0.3cm]
\end{center}
In this talk we explore the possibilities of chiral extrapolations using
the covariant formulation of chiral effective field theories (EFTs) with
baryons. In particular, a study of the convergence of the chiral
expansion and of the statistical uncertainties arising from presently
unknown  parameters in the extrapolation functions is performed for the
examples of the nucleon mass and the anomalous magnetic moment of the nucleon. In the calculation of these
quantities at next to leading one loop level, i.e. order $p^4$, we
rely on a modified scheme of infrared regularization (IR),
which will be discussed in-depth in a forthcoming publication.\\
Calculating an observable in EFT, one has a freedom
of choice which part of the analytic pieces of the result is considered to arise from loop
dynamics and which one corresponds to local operators. Taking advance of this
freedom, the IR-scheme was introduced \cite{BL} to
overcome the long-known problem of power-counting violations in the
$\overline{\textnormal{MS}}$-scheme of covariant BChPT. Compared to
standard $\overline{\textnormal{MS}}$-results, IR absorbs an infinite
sum of terms analytic in the quark mass into low
energy constants (LECs). Problems\footnote{For example the unexpanded results
can display a non-physical scale dependence or show incorrect
thresholds and singularities}
can arise due to the fact, that in
EFT at a certain order only a finite number of LECs is
included. Therefore, we propose to only
consider those analytic terms as part of short distance physics, which
can be absorbed into LECs present at the order of the calculation.\\
The quark mass dependence both for the nucleon mass and for the anomalous
magnetic moment of the nucleon calculated with this method has the
following properties: 1. A successful chiral extrapolation of currently
available lattice data to the physical point can be performed with reasonable values for the appearing parameters, i.e. values consistent with
known information from scattering theory. 2. Statistical uncertainties
are small. 
3. We do see a clear convergence of the results when going from $p^3$
to $p^4$. 4. Finally, we estimate possible effects arising at next to
leading order by using realistic quark mass dependencies for the
lowest order couplings\footnote{i.e. $g_A(m_q)$, $F_{\pi}(m_q)$ and
$m_{\pi}(m_q)$ as found in lattice simulations. This procedure
corresponds to an effective
inclusion of some fifth order effects}
and find only little impact on the best-fit curves.

\newabstract 
\begin{center}
{\large \bf Aspects of ChPT at large {\Large{\boldmath $m_s$}}}\\[0.5cm]
Juerg Gasser$^a$, {\bf Christoph Haefeli}$^{a}$, 
  Mikhail A.~Ivanov$^b$, Martin Schmid$^a$\\[0.3cm]
$^a$ Institute for Theoretical Physics, University of Bern\\
Sidlerstr. 5, 3012 Bern,  Switzerland\\[0.3cm]
$^b$ Laboratory of Theoretical Physics, Joint Institute for Nuclear
  Research\\
141980 Dubna, Moscow region, Russia\\[0.3cm]
\end{center}

\vskip 0.3cm
\noindent
Chiral perturbation theory  determines the low--energy
structure of Green functions in QCD through a systematic expansion in powers
of  quark masses and of the external momenta \cite{weinberg},\cite{GL1u2}.  
The method  has been successfully applied for
 an expansion in powers of $m_u$
and $m_d$ at fixed $m_s,m_c,\ldots$ [chiral SU(2)$\times$SU(2)] as well
 as for an
expansion in powers of $m_u$, $m_d$ and $m_s$ at fixed $m_c,m_b,m_t$ [chiral
SU(3)$\times$SU(3)]. It makes use of an effective action who's pertinent
low--energy constants (LECs) 
encode the heavy degrees of freedom that are not
explicitly contained in it. The LECs in 
SU(2)$\times$SU(2) depend on the strange
quark mass, while the ones in SU(3)$\times$SU(3)  are independent thereof. 

\noindent
If one limits the external momenta to values small compared 
to $m_s$
and treats $m_u$ and $m_d$ as small in comparison to $m_s$, the degrees of
freedom of the $K$-- and $\eta$-- mesons freeze and one can 
work out explicitly
 the strange quark mass dependence 
of the LECs in SU(2)$\times$SU(2) from the effective action 
pertaining to SU(3)$\times$SU(3). For example, for the pion 
decay constant $F$ at $m_u=m_d=0$, $m_s\neq
0$, one has at one loop \cite{GL1u2}
\begin{eqnarray}\label{F}
F = F_0\left(1 - 
\frac{m_sB_0}{32\pi^2
F_0^2}\left[\ln{\frac{m_sB_0}{\mu^2}}-256\pi^2L_4^r\right] +
\mathcal{O}(m_s^2) \right) \, .\nonumber 
\end{eqnarray}
Here, $F_0,B_0$ and
$L_4^r$ are LECs from SU(3)$\times$SU(3), and the 
renormalization scale is denoted by $\mu$.
Relations of this type generate constraints on the
possible values of the low--energy constants: information on LECs
in SU(2)$\times$SU(2) may  be translated into information 
on LECs
in SU(3)$\times$SU(3) and vice versa \cite{GL1u2}. 

\noindent
In my talk, I discussed recent 
efforts \cite{GHIS} to establish the corresponding
 relations between the LECs to two--loop accuracy.


\newabstract 
\begin{center}
{\large\bf A Chiral Two-Matrix Theory and the Chiral Lagrangian}\\[0.5cm]
{\bf Poul H. Damgaard}\\[0.3cm]
Niels Bohr Institute,
Blegdamsvej 17,
DK-2100 Copenhagen
Denmark\\[0.3cm]
\end{center}

In the $\epsilon$-regime of QCD an efficient
method for extracting the chiral condensate $\Sigma$ is based on the distributions
of Dirac operator eigenvalues. Previously, the analogous technique for 
the pion decay constant $F_{\pi}$ 
was based on introducing a vector source (``imaginary isospin chemical
potential'').
Compact analytical expressions have been found for the quenched case
\cite{DHSS} and that of two light dynamical quarks \cite{DHSST}. It is computationally
difficult to go beyond these two cases.

One
would like to find a Random Matrix Theory \cite{ShurV} that describes the inclusion
of imaginary isospin chemical potential. Because it seeks
the microscopic spectra of Dirac operators with different chemical potential for
$u$ and $d$ type quarks, it turns out to be given by a Random Two-Matrix Theory \cite{ADOS}. We solve
this theory for finite size $N$ of the matrices by means of biorthogonal polynomials.
By deriving the relevant kernels we find explicit formulas for all $(n,k)$ mixed
or unmixed spectral correlation functions of the two involved Dirac operators. In
the microscopic scaling limit we compute the corresponding analytical expressions as
well. These correspond to the leading expressions of the $\epsilon$-expansion based
on the chiral Lagrangian. In the two cases where comparison to the
chiral Lagrangian framework are possible \cite{DHSS},\cite{DHSST} we find exact agreement.
A useful by-product is the corresponding 
analytical expressions for spectral correlation functions of Dirac operators
with chemical potential, evaluated in ensembles with dynamical quarks of {\em zero}
chemical potential. These
expressions allow for a determination of the pion decay constant
by means of ordinary lattice configurations.

\newabstract 

\begin{center}
{\large\bf The {\boldmath$B\to D^* l \nu$} form factor from lattice QCD}\\[0.5cm]
\textbf{Jack Laiho}$^1$ on behalf of the MILC and Fermilab collaborations\\[0.3cm]
$^1$Theoretical Physics Department, Fermilab,
Batavia, IL 60510 \\[0.25cm]
\end{center}

\noindent
The CKM element $V_{cb}$ is important for the phenomenology of
flavor physics in determining the apex of the unitarity triangle
in the complex plane.  It is possible to determine $|V_{cb}|$ from
both inclusive and exclusive semileptonic $B$ decays, and they are
both limited by theoretical uncertainties. The inclusive method
makes use of the heavy quark expansion \cite{ball}, \cite{bigi}, but is
limited by the breakdown of local quark-hadron duality, the errors
of which are difficult to estimate.  The exclusive method requires
reducing the uncertainty of the form factor ${\cal
F}_{B\rightarrow D^*}$, which has been calculated with lattice QCD
in the quenched approximation using the double ratio method
\cite{hashimoto}. Given the phenomenological importance of this
quantity we have revisited this calculation of ${\cal
F}_{B\rightarrow D^*}$ using the 2+1 flavor MILC lattices with
improved light staggered quarks \cite{MILC}.
The lattice calculation was done on the MILC coarse lattices ($a
\approx 0.125$ fm) where the light quarks were computed with the
``asqtad" action.  The heavy quarks were computed using the clover
action with the Fermilab interpretation in terms of HQET
\cite{elkhadra}. We have looked at three light masses at the full
QCD points, $m_{valence}=m_{sea}$.  The lightest such mass was
around $m_{s}/7$.
The chiral perturbation theory (ChPT) for heavy light mesons with
a Wilson-like heavy quark and a staggered light quark was worked
out by Aubin and Bernard in \cite{aubin}, and the ChPT relevant
for $B \to D^*$ was done in \cite{laiho}.
The plan for the future is to add statistics by running on
multiple time sources, and to run on additional lattice spacings
generated by MILC and the Fermilab lattice group in order to
determine the lattice spacing dependence. We will incorporate the
matching coefficients into the analysis, and of course, we will do
a careful analysis of all the systematic errors in order to report
a final number for the $B \to D^* \ell \nu$ form factor.

\newabstract 

\begin{center}
{\large\bf Heavy Quark Effective Theory on the lattice: The b-quark
mass including O{\boldmath$(1/m_{\rm b})$} }\\[0.35cm]
{\bf M.~Della~Morte}$^1$\\[0.25cm]

\vspace{-0.1cm}
$^1$CERN, Physics Department, TH Division,
CH-1211 Geneva 23, Switzerland\\[0.25cm]
\end{center}

The b-quark is too heavy to be treated dynamically on the lattice. Indeed,
in units of nowadays affordable resolutions $am_{\rm b} >> 1$.
A theoretically attractive option is to use effective theories like
Heavy Quark Effective Theory (HQET). Introduced in~\cite{EH}, it provides 
the correct asymptotic description of QCD correlation functions in the
limit $m_{\rm b} \!\to\! \infty$. Subleading effects are described by 
higher dimensional operators whose coupling constants are formally 
O($1/m_{\rm b})$ to the appropriate power. The degrees of freedom
in the effective theory are strongly coupled and therefore a 
non-perturbative approach is needed. In addition HQET has to be matched 
to the full theory, a step, which must as well be performed 
non-perturbatively beyond leading order in $1/m_{\rm b}$. 

A framework for 
non-perturbative HQET on the lattice has been introduced in~\cite{NpHQET}.
We present here a specific application of that approach to the 
computation of the b-quark mass in the quenched approximation. 
The details can be found in~\cite{mb1}.
In a first step we match HQET and QCD in a small volume ($L=0.4$ fm).
To make contact with phenomenology the HQET expressions of the relevant 
quantities are then  evolved to large volume. The essential tools in this
step are the step scaling functions, which we introduce to describe the
effects of a change in the linear size $L$ of the system  by a factor two.
Eventually the b-quark mass is  fixed using the (spin averaged) 
$B_{\rm s}$ meson mass. We include the subleading, $1/m_{\rm b}$, 
terms in each step. Our final result is $\overline{m}_{\rm b}
(\overline{m}_{\rm b})=4.347(48)$ GeV in the $\overline{\rm MS}$ scheme.
We have implemented twelve different matching conditions for
this computation. Although the contributions at a given order may differ
in the twelve different cases, once the LO and the NLO terms are summed 
together all the estimates agree. That implicitly suggests that the 
$1/m_{\rm b}^2$ corrections to the quark mass are extremely small as expected.
We plan to extend the method to the computation of the decay constant
$F_{\rm B_{\rm s}}$, which we have already computed in the static 
approximation~\cite{Fb} and to real (unquenched) QCD.

\newabstract 

\begin{center}
{\large\bf Overlap hypercube quarks in the {\boldmath$p$}- and the
{\boldmath$\epsilon$}-regime}\\[0.5cm]
{\bf Wolfgang Bietenholz}$^1$, Stanislav Shcheredin$^2$
and Jan Volkholz$^1$\\[0.3cm]
$^1$ Institut f\"{u}r Physik, Humboldt Universit\"{a}t zu Berlin,\\
Newtonstr.\ 15, D-12489 Berlin, Germany \\[0.3cm]
$^2$ Fakult\"{a}t f\"{u}r Physik, Universit\"{a}t Bielefeld \\
D-33615 Bielefeld, Germany \\

\end{center}

The overlap hypercube fermion is a variant of a chirally 
symmetric lattice fermion, which has a much higher degree
of locality than the standard overlap fermion. 
We applied this formulation in quenched QCD simulations
with light quarks \cite{NPB}, which were performed with HLRN
machines on a $12^{3} \times 24$ lattice
at $\beta = 5.85$ (corresponding to a lattice spacing
of $a \simeq 0.123~{\rm fm}$ according to the Sommer scale).
The topological susceptibility that we obtained
supports the Witten-Veneziano scenario to explain 
the mass of $\eta '$.

In the $p$-regime we evaluated 
the masses of light pseudoscalar and vector mesons. Although
we reach $m_{\pi} < 300 ~{\rm MeV}$ it is difficult to extrapolate
the corresponding value for $F_{\pi}$ --- measured from a matrix
element --- to the physical point. However, we found a stable
axial current renormalisation constant $Z_{A}$ close to $1$,
in contrast to the results with the standard overlap operator.

The use of tiny bare quark masses in the same volume took us to
the $\epsilon$-regime, where we evaluated the leading Low Energy Constants 
$\Sigma$ and $F_{\pi}$ with various methods \cite{NPB}.
The densities of low lying Dirac eigenvalues were fitted to
predictions by chiral Random Matrix Theory, which yields
$\Sigma \approx (300 ~{\rm MeV})^{3}$.
Next we compared the axial current correlators to formulae of
quenched Chiral Perturbation Theory,
which led to $F_{\pi} \approx 108 ~ {\rm MeV}$.
On the other hand, a method using only the zero mode contributions
to the pseudoscalar correlators gave a lower value, close to
the phenomenological $F_{\pi}$. We conclude that these methods have
the potential to determine Low Energy Constants of the chiral Lagrangian
directly from the first principles of QCD, but precise values have to
await the feasibility of simulations with dynamical, chiral quarks.

As a step into this direction, we simulated dynamical overlap
hypercube fermions in the Schwinger model using a Hybrid Monte Carlo 
algorithm with a simplified force, 
where we tested algorithmic requirements like 
reversibility \cite{Lat06}.
The measurements confirm an extremely high level of locality.
We also evaluated the chiral condensate
based on the Dirac spectrum in the $\epsilon$-regime, and we
found good agreement with low energy predictions at two very light
fermion masses.

\newabstract 

\begin{center}
{\large\bf {\boldmath$\sigma(600)$} as a Tetraquark Mesonium and 
the Pattern of Scalar Mesons}\\[0.5cm]
{\bf Keh-Fei Liu}\\[0.3cm]
Dept. of Physics and Astronomy, University of Kentucky,\\
Lexington, KY 40506, USA\\[0.3cm]
\end{center}

    Scalar mesons are not as well known as the pseudoscalar and vector mesons 
in terms of their
$SU(3)$ classification, their particle content and their spectrum. 
It has been known  
that there are too many experimental states for the $q\bar{q}$ nonet, 
possibly by a factor of two.
We shall use lattice QCD calculations and combine with phenomenology to 
help understand the
nature of the scalar mesons, their particle content and their $SU(3)$ 
classification. 

    A recent lattice calculation~\cite{mathur} with the overlap 
fermion on $16^3 \times 28$ and
$12^3 \times 28$ quenched lattices with the Iwasski gauge action at $a= 0.2$ fm has
shown that the isovector scalar mesons with the $\bar{\psi}\psi$ interpolation field 
has an unusual behavior as it approaches the chiral limit. Below the strange
quark mass, its mass 
becomes flat and is almost independent of the quark mass, in contrast to 
other hadrons. This 
calculation which reaches a pion mass as low as 180 MeV is consistent with
earlier 
findings with heavier quark masses~\cite{mathur}.
Its chiral limit extrapolation indicates that $a_0(1450)$ is the $q\bar{q}$
meson 
and the $a_0(980)$ 
below it is not seen with the $\bar{\psi}\psi$ interpolation field. 
It explains the experimental fact
that $K_0^*(1430)$ is basically degenerate with $a_0(1450)$. From the near 
degeneracy of
$a_0(1450)$, $K_0^{*}(1430)$, and $f_0(1500)$, we build a model for 
the mixing of 
$f_0(1370), f_0(1500)$, and $f_0(1710)$ through the annihilation of 
$u\bar{u}, d\bar{d}$ and
$s\bar{s}$ with a slight $SU(3)$ breaking in their mixing and
 decays~\cite{ccl}. It is found
that ratios of pseudoscalar meson decays can be well descried by this 
simple model. The interesting
thing is that we found $f_0(1500)$ is almost a pure octet, $f_0(1710)$ 
an almost pure glueball,
and $f_0(1370)$ is mainly a singlet with $\sim 10\%$ glueball content. 

    We also calculated tetraquark mesonium with I=0 $\pi\pi$ interpolation 
field and
found a state around 550 MeV in addition to the $\pi\pi$ scattering 
states in the range where
pion mass is lower than 250 MeV. Using the volume study of their spectral 
weight~\cite{mathur_roper},\cite{mathur_penta}, we verified that the lowest state 
around 300 MeV is
the interacting $\pi\pi$ state and the state at $\sim 550$ MeV is a 
one-particle state which
we believe is the $\sigma(600)$. Our present lattice calculation only 
confirms its existence.
To get its precise mass and width, one needs to vary the lattice size to bring down the
scattering state with one unit of lattice momentum to mix with it and observe how far the
levels avoid each other. 

    Based on the above two lattice calculations, we speculate on the 
following pattern for 
scalar mesons: $a_0(980), f_0(980)$ 
and $\kappa(800)$ form a tetraquark mesonium octet with the $\sigma(600)$ as the singlet.
On the other hand, $a_0(1450), K_0^*(1430)$, and $f_0(1500)$ form a $q\bar{q}$ octet with
$f_0(1370)$ as its singlet and a almost pure glueball which is $f_0(1710)$. This needs
additional experimental and lattice confirmation.

    This work is partially supported by the US DOE grant DE-FG05-84ER40154. I thank the
organizers of the workshop on Lattice QCD, Chiral Perturbation Theory, and 
Hadron Phenomenology
for giving me a chance to present this work.

\newabstract 

\begin{center}
{\large\bf Moments of Generalized Parton Distribution Functions}\\[0.5cm]
{\bf Marina Dorati}$^1$, Thomas R. Hemmert$^2$\\[0.3cm]
$^1$Dep. Theor. Phys., Univ. of Pavia, 27100 Pavia, Italy\\[0.3cm]
$^2$Physik Dep. T39, TU M{\"u}nchen, D-85747 Garching\\[0.3cm]
\end{center}

We apply the formalism of covariant Baryon Chiral Perturbation Theory (BChPT) to the analysis of the moments of the
Generalized Parton Distributions (GPDs) of the nucleon. We concentrate on the isovector $n=1$ case and perfom a $\mathcal{O}(p^4)$ calculation to obtain the Generalized Form Factors $A_{2,0}^{v}(q^2)$, $B_{2,0}^{v}(q^2)$ and $C_{2,0}^{v}(q^2)$. These results can be used for chiral extrapolations of lattice data for GPDs.
Of particular interest is the forward limit, where $A_{2,0}^{v}(q^2=0)$ reduces to the averaged momentum fraction $\langle x\rangle_{u-d}$, a quantity known from phenomenology. Lattice data are also available for this moment. The figure shows two different chiral extrapolation functions. The dashed one indicates the influence of the well known leading chiral logarithm \cite{JS}, unable to reach the plateau observed in recent lattice simulations \cite{LAT}. The full curve shows our $\mathcal{O}(p^4)$ covariant BChPT result in the modified Infrared Regularization scheme \cite{GH}. We conclude \cite{DH} that a whole tower of quark mass dependent terms is required in order to understand $\langle x\rangle_{u-d}(m_{\pi})$, in contrast to other recent claims \cite{Richards}.

\begin{figure}[h]
\centering
\includegraphics[width=.6\textwidth]{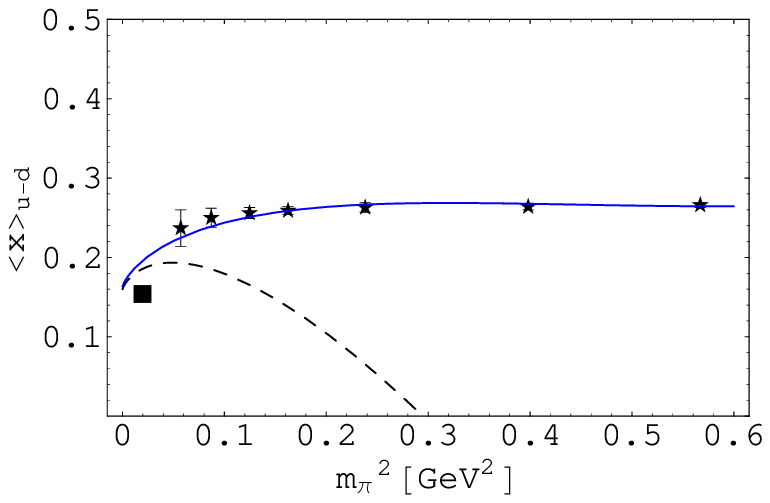}
\end{figure}

\newabstract 

\begin{center}
{\large\bf ChPT for Generalized Parton Distributions (GPDs)}\\[0.5cm]
M. Diehl$^1$, A.~Manashov$^2$ and {\bf A.~Sch\"afer}$^2$\\[0.3cm]
$^1$Deutsches Elektron-Synchroton, DESY, 22603 Hamburg\\[0.3cm]
$^2$Institute for Theoretical Physics, University of Regensburg, 
93040 Regensburg\\[0.3cm]
\end{center}

Generalized parton distributions provide a unified parametrization of 
hadron structure. They allow one to combine in the most efficent way 
the information obtained from various inclusive and exclusive reactions
and permit to extract informations 
on the internal hadron structure which cannot 
be obtained directly by experi\-ment. However, the extraction of GPDs
from experiment is very non-trivial, especially to NLO and NNLO
in $\alpha_s$,
which is needed in view of the typical lowish $Q^2$. Therefore, 
lattice calculations of moments of GPDs on the lattice are especially 
important for GPD physics \cite{r1}. The extrapolation of such results
to physical quark masses requires a ChPT analysis of these moments.\\ 
We performed such an analysis, both for the pion GPDs \cite{r2} 
and for the isosinglet nucleon GPDs \cite{r3}. 
We proceed by first constructing 
the leading order ChPT equivalents of the operators apearing in the 
OPE analysis of moments of GPDs. We then construct the next-to-leading 
contributions and calculate the 1-loop corrections to the leading ones.\\
The results allow one to identify the leading analytic terms which apear and 
permit to relate some LECs to known physics.

\newabstract 

\begin{center}
{\large\bf The QCD vacuum as seen by overlap fermions}\\[0.5cm]
{\bf Volker Weinberg}\\
Deutsches Elektronen-Synchrotron DESY, 15738 Zeuthen, Germany \\[0.3cm]

-- QCDSF / DIK Collaboration -- \\[0.4cm]
\end{center}

With the advent of overlap fermions, which are the cleanest known theoretical 
description of lattice fermions implementing chiral symmetry, it has recently 
become possible~\footnote{See Ref.~\cite{QCDSF} and references therein.} 
to analyze the vacuum structure of Yang-Mills theory and the nature of the
chiral phase transition of QCD from first principles using a fermionic approach. 

We report results based on  $\cal O$(150) low lying overlap eigenmodes on 
quenched L\"uscher-Weisz configurations on various T=0 lattices that were 
obtained within the QCDSF-collaboration \cite{QCDSF}.
The analysis is also extended to
dynamical finite temperature configurations generated 
by the DIK-collaboration \cite{DIK} using the
clover-improved Wilson action.

The distribution of the lowest modes in selected topological sectors and 
the spectral density  is in good agreement with the finite-volume
prediction of quench\-ed chiral perturbation theory and Random Matrix Theory. 
In the vicinity of the finite-T phase transition the 
emergence of a gap in the spectrum is observed.

Truncating the spectral decomposition of both the overlap-based 
topological charge density operator and the gluonic field strength tensor allows for a 
removal of fluctuations at the scale of the cutoff.
Comparing the UV-filtered operators, we observe a strong correlation between 
the peaks of the topological charge density, the regions of high (anti-)selfduality  
and the domains of strong local chirality of the lowest localized eigenmodes. 
In the chiral-symmetry restored phase, on the other hand, the signal of local 
chirality and (anti-)selfduality is strongly suppressed. 
In contrast to the singular low-dimensional structure of the untruncated  
topological charge density, the UV-filtering reveals a structure of the 
T=0 vacuum much more reminiscent of semiclassical-like instanton pictures.

\newabstract 

\begin{center}
{\large\bf The Chiral Regime of QCD in the Instanton Liquid Model}\\[0.5cm]
Marco Cristoforetti$^1$, {\bf Pietro Faccioli}$^1$, Marco C. Traini$^1$, 
and John W. Negele$^2$\\[0.3cm]
$^1$ Dipartimento di Fisica and I.N.F.N., Universit\`a degli Studi di Trento,\\
 Via Sommarive 15, Povo (Trento) 38050 Italy.\\[0.3cm]
$^2$ Center for Theoretical Physics
Massachusetts Institute of Technology,\\
NE25-4079, 
77 Massachusetts Ave, Cambridge, MA 02139-4307, USA.\\[0.3cm]
\end{center}
Contemporary lattice QCD calculations have highlighted the need to understand 
the dependence of QCD observables on the quark mass, in order to provide
reliable extrapolation formulas. In addition to this motivation, studying the 
microscopic dynamical mechanisms involved in the transition into the chiral 
regime of QCD may shed light on the structure of the non-perturbative
 quark-quark interaction. 

In this work we  investigate the non-perturbative quark-gluon dynamics
involved  in such a transition in the context of the Interacting Instanton 
Liquid Model (IILM) \cite{cristof}. 
By computing the nucleon and pion masses for a wide range of quark masses, 
we show that the IILM reproduces the existing lattice data for pion masses 
in the range  $\simeq~500 - 600$~MeV. 
Fitting the nucleon masses, using covariant Baryon $\chi$PT at order 
$\mathcal{O}(p^4)$, we obtain the chiral coefficients in good agreement 
phenomenology. In particular the parameters $c_1$ and $c_3$ match well 
the values extracted from the analysis of the existing $\pi N$ scattering data. 
On the other hand, the same chiral extrapolation performed using the 
available MILC data instead of the IILM data leads to chiral 
parameters $c_1$ and $c_3$ which are inconsistent with the experimental data. 
Such an inconsistency is probably  due to the fact that the available 
MILC data have not been extrapolated to the continuum limit. 
We also show that the IILM agrees well also with Chiral Perturbation 
Theory in the small quark mass regime. For example, we found that 
the spectral density of the Dirac operator calculated in the IILM agrees 
with the behavior expected from chiral perturbation theory for two flavors.
Hence, we can argue that the IILM can be used to study the transition into 
the chiral regime of QCD.
We identify a characteristic quark energy scale,  
$m^\star$ = 80 MeV, which determines the boundary of the near-zero mode 
zone of the quark propagator in the instanton background. We argue that 
this scale governs the transition into the chiral regime and we discuss 
its physical interpretation.

\newabstract 

\begin{center}
{\large\bf Discussion: Lattice meets Chiral Perturbation Theory}\\[0.5cm]
{\bf Hartmut Wittig}$^1$ (convenor)\\[0.3cm]
$^1$Institut f\"ur Kernphysik, Universit\"at Mainz, D-55099 Mainz,
Germany.\\[0.3cm] 
\end{center}

I summarise the main points of the discussion held at the workshop:\\[0.3cm] 
{\bf What is the status of lattice determinations of LECs?}\\
Even ``simple'' LECs such as $F_\pi$ are difficult to extract,
signified by the large variation of results for this quantity. The
compatibility of results from different simulations must be better
understood, and an attempt to do this using recent data was presented
\cite{scorzato}.

It was noted that the precision with which lattice determinations of
LECs are quoted by some groups, is far too high, given the quality of
the lattice data.

\bigskip\par\noindent
{\bf Have we made contact between ChPT and lattice QCD?}\\
Often one finds that authors regard the presence of chiral logs in
lattice data as the criterion for contact between lattice QCD and
ChPT, since the coefficient of the chiral log is universal. It was
pointed out \cite{obaer} that this is not necessarily true, since this
coefficient acquires regularisation-dependent contributions if chiral
symmetry is not respected.

In another contribution \cite{ugm} the two-loop analysis for the quark
mass dependence of $F_\pi$ \cite{ColDurr} was compared to lattice data
by NPLQCD \cite{Beane:2005rj} with pion masses as low as
300\,MeV. While the two-loop formula describes the lattice data
reasonably well, the claimed precision for the lattice determination
of $F_\pi$ seemed exaggerated.

\bigskip\par\noindent
{\bf What do ChPT people need from the lattice?}\\
More lattice data at smaller pion masses ($m_\pi < 300$\,MeV) must
become available for more reliable determinations of LECs. In mesonic
ChPT, the $O(p^4)$ LEC $L_6$ is not well constrained by phenomenology
and should be a target for lattice calculations. Also, there are few
results for LECs in the baryonic sector. Here, the lattice should try
to find ways to determine the so-called class~II LECs, or symmetry
breakers, since class~I LECs having a dynamical origin (i.e. those
multiplying terms with derivatives) are more easily determined from
phenomenology \cite{ugm_lat05}.

It was suggested that a compilation of the status of determining LECs
be made available, possibly in the form of a web-page. Some
information can also be found in \cite{ugm_lat05}.

\bigskip\par\noindent
{\bf What do lattice people need from lattice people?}\\
A standardised way of quoting results in order to facilitate
comparison of results from several groups \cite{gsch}. In particular,
one standard choice of lattice scale (e.g. $r_0$) should be employed
in addition to any other scale that collaboration might favour
otherwise.


\end{document}